\documentclass[aps,pra,twocolumn,showpacs,superscriptaddress]{revtex4}
\usepackage{hyperref}
\usepackage{amssymb}
\usepackage{amsmath}
\usepackage{epsfig}

\begin{document}
\title{Coulomb Rescattering in Nondipole Interaction of Atoms with  Intense Laser Fields}

\author{J. F. Tao}
\email[]{These authors contributed equally to this work.}
\affiliation{National Laboratory of Science and Technology on Computational Physics, Institute of Applied Physics and Computational Mathematics, Beijing 100088, China}
\author{Q. Z. Xia}
\email[]{These authors contributed equally to this work.}
\affiliation{National Laboratory of Science and Technology on Computational Physics, Institute of Applied Physics and Computational Mathematics, Beijing 100088, China}
\author{J. Cai}
\affiliation{School of Physics and Electronic Engineering, Jiangsu Normal University, Xuzhou 221116, China}
\author{L. B. Fu}
\affiliation{National Laboratory of Science and Technology on Computational Physics, Institute of Applied Physics and Computational Mathematics, Beijing 100088, China}
\affiliation{CAPT, HEDPS, and IFSA Collaborative Innovation Center of MoE, Peking University, Beijing 100871, China}
\author{J. Liu}
\email[]{liu\_jie@iapcm.ac.cn}
\affiliation{National Laboratory of Science and Technology on Computational Physics, Institute of Applied Physics and Computational Mathematics, Beijing 100088, China}
\affiliation{CAPT, HEDPS, and IFSA Collaborative Innovation Center of MoE, Peking University, Beijing 100871, China}
\affiliation{Center for Fusion Energy Science and Technology, China Academy of Engineering Physics, Beijing 100088, China}

\begin{abstract}
{We investigate the ionization dynamics of atoms irradiated by an intense laser field using a semiclassical model that includes magnetic Lorentz force in the rescattering process. We find that, the electrons tunneled with different initial transverse momenta (i.e., perpendicular to the instantaneous electric field direction) distributed on a specific circle in the momentum plane can finally converge to the same transverse momentum after experiencing Coulomb forward scattering. These electron trajectories lead to a bright spot structure in the 2D transverse momentum distribution, and particularly in the long-wavelength limit, a nonzero momentum peak in the direction antiparallel to the laser propagation (or radiation pressure) direction. Making analysis of the subcycle dynamics of rescattering trajectories, we unveil the underlying mechanism of the anti-intuitive peak. Beyond the strong field approximation and the dipole approximation, we quantitatively predict the spot center and the peak position. Our results are compared with a recent experiment and some theoretical predictions are given.}
\end{abstract}
\pacs{32.80.Rm, 31.15.xg, 32.80.Fb}

\maketitle

\textbf{Introduction}
Rescattering or recollision, i.e.,  a released electron
collides under intense laser forces with the core, is at
the heart of strong-field phenomena\cite{recollision,book_JieLiu,Review_Becker} and dramatically affects above threshold
ionization (ATI) spectra\cite{ati,plateau}, high-order harmonic generation (HHG)\cite{hhg}, double ionization\cite{book_yedifa}, and neutral atom acceleration \cite{Eichmann2009, xia}, etc.
Most previous studies are based on a dipole approximation, that is, the laser magnetic component is neglected in the description of Coulomb recsattering \cite{DP}. It simplifies problems but leaves out a substantial property of traveling electromagnetic waves, namely, the radiation pressure that  exerts on the scattered charged particles.

Experimental and theoretical analyses on photon momentum partition in atomic ionization have been conducted that reveal profound characteristics of the radiation pressure\cite{smeenk,tau}. Particularly in the long wavelength limit where the electron's excursion in the laser propagation direction caused by the magnetic  Lorentz force is comparable to the atomic length scale (Bohr radius $r_B$), i.e., $U_p/2m_e\omega c \sim r_B$ with $U_p$ the ponderomotive potential, $m_e$ the electron mass and $c$ the vacuum light speed, the dipole approximation breaks down and the genuine electromagnetic vector potential with spatial dependence should be considered \cite{Reiss}. The extra factor complicates the study of Coulomb effects in laser-atom interaction but leads to very intriguing results.
For instance, in an experiment using a laser wavelength of 3400 nm, a striking peak in the transverse momentum distribution is found to shift towards negative values on the laser beam propagation axis\cite{ludwig}, which seems to contradict the prediction of positive transverse momentum shifts from an intuitive picture of radiation pressure effect\cite{tau, corkum2015}. This anti-intuitive experimental observation might be due to the interplay between nondipole interaction and Coulomb rescattering and urgently calls for insightful investigation from the theoretical side.

\begin{figure*}[t]
\includegraphics[width=0.79\linewidth]{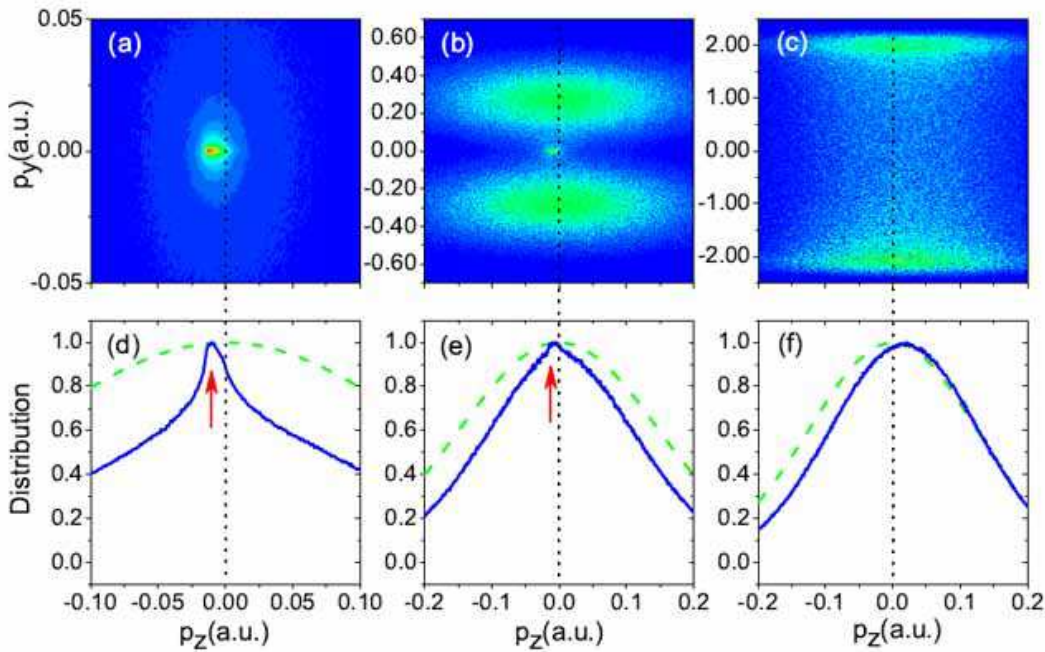}
\caption{(Color online) The electron momentum distributions of Xe atom on the $p_y$-$p_z$ plane (upper row) and the momentum distributions along the propagation direction (lower row). The laser parameters are: (a) and (d), 3400 nm, ellipticity $\rho=0.0$(LP); (b) and (e), 3400 nm, $\rho=0.1$; (c) and (f), 3400 nm, $\rho=1.0$(CP). Laser intensity is the same for all panels, 0.06 PW$/cm^2$. The distributions of the initial transverse momentum after tunneling are plotted in dashed green.}
\end{figure*}

In this paper, we study the ionization dynamics of xenon atoms irradiated by intense long wavelength propagating laser fields addressing the Coulomb effects during photoelectron rescattering, beyond the traditional strong field approximation (SFA)\cite{KFR} and dipole approximation.
Interestingly, we find that the electrons tunneled with different initial transverse momenta distributed on a specific circle in the momentum plane can converge to the same final transverse momentum after experiencing Coulomb forward scattering. It gives rise to a bright spot structure in the 2D transverse momentum distribution  and particularly a nonzero momentum peak in the direction antiparallel to laser propagation in the long-wavelength limit.
Our theory accounts for recent experimental observations\cite{ludwig} and resolves the standing controversy.

\textbf{Model calculation}
We model the strong-field tunneling ionization process from a classical trajectory perspective \cite{book_JieLiu}. In our model, the initial position of the electron along the instantaneous electric field direction is derived from the
Landau effective potential theory\cite{landau}. The spreading of the tunneled electron wave packet is described by a
Gaussian-like transverse velocity distribution\cite{delone}. Each electron
trajectory is weighted by the ADK tunneling ionization rate. The initial longitudinal momentum is set to be zero.
 After tunneling, the electron evolution in the combined
oscillating laser field and Coulomb potential is solved via the
Newtonian equation\cite{hu}. The laser field is described by a plane electromagnetic wave (EMW) propagating in the positive Z direction,
 and its vector potential is $\textbf{A}(\textbf{r},t)  = \epsilon_0 e^{\frac{-4 {\ln2} }{(8\pi)^2}\eta^2}(-  \sin\eta \ \hat{x} + \rho  \cos\eta \ \hat{y})/\omega\sqrt{1+\rho^2}$, where $\eta$ is the Lorentz invariant phase $\eta = \omega t - k z$, $k = \omega/c$ is the wave vector, $\epsilon_0$ is the maximum field strength, and $\rho$ is the ellipticity.
 The corresponding electric field component $\textbf{E}(\textbf{r},t)$ and magnetic field component $\textbf{B}(\textbf{r},t)$ can be derived from $\textbf{A}(\textbf{r},t)$. In our model, the Lorentz force is included in the Newtonian equation after tunneling, while the influence of the magnetic field on the tunneling process \cite{smeenk, keitel} is ignored, because it is too small for the laser field strength of the current experiment as mentioned in Ref.\ \cite{ludwig}. Note that unless stated otherwise, atomic units are used throughout the paper.

 We calculate the tunnel ionization of xenon atoms in 3400nm laser with various laser parameters to exhibit the photoelectron transverse momentum distribution. The 2D  momentum distributions projected to the YOZ plane perpendicular to the major polarization X axis are plotted in Fig. 1 (a) to (c), with laser ellipticity $\rho=$0, 0.1 and 1.0, respectively. The corresponding  1D distributions along the axis of laser propagation (Z) by integrating the 2D spectra over Y axis are presented in (d) to (f), respectively.

We observe a bifurcation in the 2D momentum distribution from one spot to a dumbbell shape from LP in Fig. 1(a) to circularly
polarized (CP) field in (c),
which can be simply interpreted by the simple man model as a result of the drift velocity \cite{Landsman2013, limin_2013}.
In the case of LP field, a small spot is seen near the center of the spectral distribution. For the elliptically polarized (EP) field, the transverse drift velocity of the tunneled electron increases so that the central spot becomes vaguer as seen in Fig. 1 (b). In the CP field, the tunneled electron has a negligible probability to revisit its parent ion \cite{fu}, no apparent focusing spot is seen in the central area of the 2D spectral distribution and the 1D distribution is of a gaussian-like shape.

 Note that, in Fig. 1(a), the position of the bright spot shifts to the negative Z direction which is reverse to the direction of the laser radiation pressure caused by the intrinsic  magnetic field Lorentz force \cite{smeenk,tau}. Correspondingly, a prominent peak structure shift is seen in the integrated 1D momentum distribution of Fig. 1 (d), which is also observed in the recent experiment\cite{ludwig}. However, according to our statistics, although the peak of the distribution is negative, the average value $\bar{p}_z$ over the total distribution still keeps positive, i.e., the total momentum gain is along the radiation pressure direction. This result is also consistent with previous momentum partition calculations\cite{tau,corkum2015}.
 This radiation pressure effect is more obvious in circular polarization case (Fig. 1(f)) where the pressure makes the total distribution profile move towards the laser propagation direction.

\begin{figure}
\includegraphics[width=0.9\linewidth]{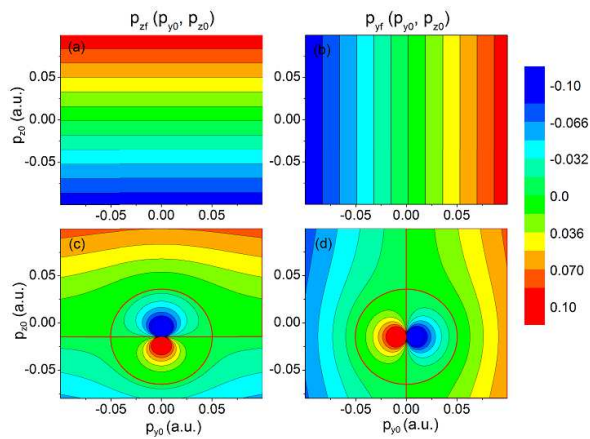}
\caption{(Color online) Contour plots of the asymptotic momentum $P_{zf}$ (a and c) and $P_{yf}$ (b and d) versus the initial transverse momentum at a fixed tunneled phase $\omega t_0=0.3$ for 3400 nm LP field. Panels (a) and (b) are without Coulomb potential while panels (c) and (d) are with Coulomb potential.}
\end{figure}

\textbf{Saddle point and the subcycle dynamics}
 To further scrutinize the origin of the bright spot and the peak structure in the 2D and 1D distributions, contour plots of the deflection function $p_{zf}(p_{y0},p_{z0})$ and $p_{yf}(p_{y0},p_{z0})$ are illustrated in Fig. 2, with a fixed tunneling phase $\omega t_0 = 0.3 \ a.u.$ for simplicity, to ensure only one return occurs.
Figure 2 (a)(b) represent the contour plots with the Coulomb potential removed artificially.
 In contrast to the straight lines in (a)(b), the contour plots in Fig. 2 (c)(d) can be divided
 into two regimes according to  the strength of the Coulomb effect:
 The contour lines are distorted severely in the strong deflection regime which is due to head-on recollisions;
 nearly straight curves in the weak deflection regime where the electron is slightly scattered by the Coulomb potential.
 Overall translation of the contour line towards the negative $p_z$ side is the result of radiation pressure.
 More interestingly, there is a circle-shape contour line that separates the above two regimes. The  electron trajectories  launched from the circle will experience forward scattering and finally converge to the same final transverse momentum. 
 The electron orbits around the circle are responsible to the bright spot and peak structure.

 Our above discussions are based on the classical trajectory perspective\cite{book_JieLiu} which can provide clear physical pictures. These orbits may also have quantum implications since the interference effects are important in laser-atom interaction physics: electrons with different tunneling moments generate inter-cycle and intra-cycle interference patterns which form and modulate the discrete multiphoton peaks in an ATI energy distribution\cite{arbo}, and the holography pattern generated by photoelectron is claimed to be the interference between the directly ionized wavepacket and the rescattering one\cite{huismans}.
 It will be an interesting to treat the effect with first-principles quantum-orbit picture\cite{Pisanty2016}.

\begin{figure}
\includegraphics[width=0.9\linewidth]{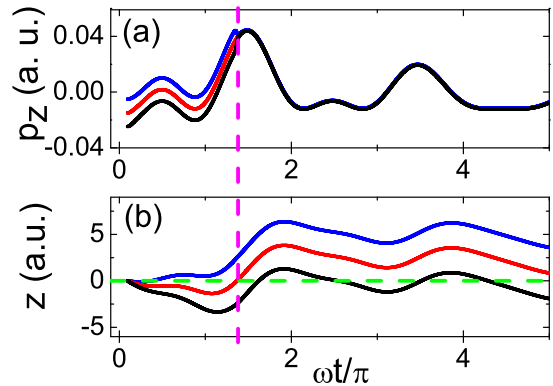}
\caption{(Color online) The evolution of the electron trajectories originating near the saddle point: (a)the evolution of $p_z$. (b) the motion along the laser propagation direction.  The vertical dashed purple line throughout(a)(b) indicates the moment when the electron returns, i.e., $x(\eta_r)=0$.}
\end{figure}

On the other hand,  another horizontal contour line intersects the circle, yielding two saddle points in the deflection function where $\partial p_{zf}/\partial p_{z0} =0$ and $\partial p_{zf}/\partial p_{y0} =0$. The saddle points can lead to accumulation and produce singularities \cite{hove} in the final momentum distribution as demonstrated in Fig. 1 (d) and (e). We illustrate in Fig. 3(a)(b) the evolution of photoelectron trajectories in the vicinity of the saddle point(with fixed $p_{y0}$ and initial phase $\eta_0$).

In contrast to the saddle-point structure in the bunching of longitudinal momentum in linearly-polarized laser for the soft recollision\cite{rost},  the saddle points here are responsible for electron accumulation in the direction perpendicular to the laser field polarization\cite{ChengpuLiu, Shafir, Wolter}. The asymmetric line-shape of the peak structures in Fig.\ 2(d) is due to the high dimension effect\cite{hove} that the final $p_{zf}$ is a function of the tunneling 2D transverse momenta as well as the initial phase.
Differently, with the dipole approximation, the peak locates exactly at zero and the corresponding line-shape is symmetric\cite{comtois,rudenko,dura}.

\textbf{Analytical derivation and comparison with experiment}
We now attempt to estimate the spot center and the peak position quantitatively according to the saddle point trajectory.
From Fig.\ 3(a), we see that the main contribution of Coulomb effect takes place at rescattering, in which the saddle point trajectory (red line) is the least altered one while the momenta of nearby trajectories are changed strongly by the Coulomb force. Based on this observation, we propose that the saddle-point trajectory satisfies
\begin{eqnarray}
z(\eta_r) \approx 0. \label{eq1}
\end{eqnarray}
Here $\eta_r(=\omega t_r)$ represents the revisiting phase(time, $t_r$) when $x(\eta_r)=0$. The saddle-point condition is verified by the evolution of displacement along Z axis in Fig.\ 3(b).

 \begin{figure}[t]
\includegraphics[width=0.9\linewidth]{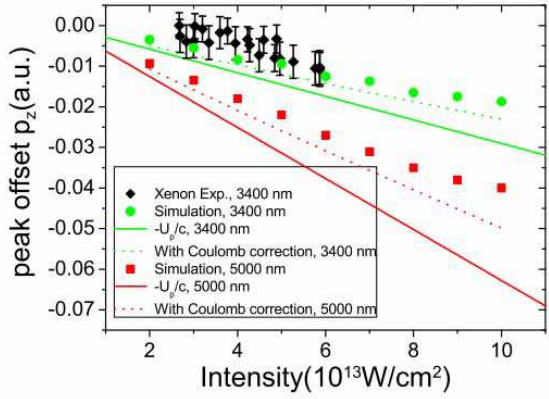}
\caption{(Color online)  The peak shift of $p_z$ distribution versus the laser intensity. The experimental results with 3400 nm LP laser are plotted with black diamonds with error bars. Numerical simulation results with 3400nm and 5000 nm laser are plotted with green and red circles, respectively. The analytical formula with or  without Coulomb correction are plotted with dashed or solid lines. }
\end{figure}

In rescattering, the transverse momentum of the tunnelled electron is influenced by the laser field as well as the Coulomb potential, so that the final momentum in the laser propagation direction can be written as,
\begin{eqnarray}
p_{zf}=p_{z0}+\Delta p_{F} +\Delta p_{zc}. \label{eq2}
\end{eqnarray}
Here, $p_{z0}=-\frac{\int_{\eta_0}^{\eta_r}d\eta(\textbf{A}(\eta)-\textbf{A}(\eta_0))^2}{2c(\eta_r-\eta_0)}$  is the initial transverse momentum satisfying the saddle point condition Eq. (\ref{eq1})\cite{sarachik};  $\Delta p_{F}=\frac{(\textbf{A}(\infty)-\textbf{A}(\eta_0))^2}{2c}$  is the impulse by the laser field along $Z$  direction; $\Delta p_{zc}$ is the Coulomb effect. The expression of the focused momentum $p_{zf}$ in Eq. (\ref{eq2}) clearly demonstrates its dependence on the tunneled phase $\eta_0$ and relates itself to the time-resolved photoelectron holography\cite{corkum2015}.
Similarly, for the momentum along the minor polarization direction in the EP field, we have
\begin{eqnarray}
p_{yf}=&&-\frac{\int_{\eta_0}^{\eta_r}d\eta (A_y(\eta)-A_y(\eta_0))}{\eta_r-\eta_0}\nonumber\\
&&+(A_y(\infty)-A_y(\eta_0))+\Delta p_{yc}\label{eq3}
\end{eqnarray}

In order to calculate the Coulomb effect, let us recall the trajectory feature in Fig.\ 3(a) that the significant Coulomb attraction on the saddle-point trajectory happens only at the tunneling moment when the tunnelled electron is close to its parent ion. We then adopt a similar methodology as in Ref.\ \cite{goreslavski} and deduce the Coulomb attraction correction by integrating the Coulomb force along the straight trajectory driven by the instantaneous electric field. That is,  $\Delta p_{zc} \approx -\frac{1}{\omega^2} \int_{\eta_0}^\infty \frac{p_{z0}(\eta-\eta_0)}{((x(\eta_0)-E_x(\eta_0)(\eta-\eta_0)^2/(2\omega^2))^2+p_{z0}^2(\eta-\eta_0)^2/\omega^2)^{3/2}}d\eta$, and $\Delta p_{yc}$ can be expressed in a similar way.

 Now we compare our theory with experimental results. Because a bound electron tunnels most probably when the instantaneous field reaches its maximum, the peak position of the total transverse momentum distribution can be estimated by the electron trajectory tunneled at the maximum field, i.e., $\eta_0 \approx 0$. Therefore, the traveling time can be approximated as an optical period, resulting in a peak shift of $p_{zf} \approx -U_p/c$ (solid lines in Fig. 4) when ignoring $\Delta p_{zc}$.
 This simple expression predicts the overall trend with varying laser intensity but it overestimates the value of peak shift. And it also explains the observation that the peak shift is not sensitive to the initial distribution of transverse momentum after tunneling in the simulation\cite{ludwig}, because the leading terms in Eqn.\ (2), i.e., the first two terms do not change when the initial transverse momentum distribution is modified. After considering the Coulomb correction our analytic expression is consistent with the numerical calculation as well as the experimental results of  3400 nm. We have further calculated the transverse momentum distribution for 5000 nm laser fields and the theoretical predictions for the peak shift also agree with our numerical simulations.

In summary,
 we have theoretically investigated the photoelectron transverse momentum distribution of xenon atoms irradiated by an intense laser in the long wavelength limit and beyond the dipole approximation. By inspecting the subcycle dynamics of rescattering electrons, we ascribe the anti-intuitive negative shift of the peak structure in the transverse momentum distribution to the interplay of the Coulomb effect and radiation pressure.
 Our theory accounts for recent experimental observations and resolves the standing controversy on the peak structure.
 Our discussion is mainly based on a classical trajectory perspective which provides clear physical pictures.
 Our findings have important implications in the quantum aspect as well. For instance, it is of  interest to study the interplay of the effect with other interference pattern such as photoelectron holography\cite{Yurchenko2004, huismans}. That may provide a new control knob for future investigations on the Coulomb effect during electron rescattering and may be useful in manipulating strong-field atomic processes.

\section*{Acknowledgement}
This work is supported by the NFRP (2011CB921503)
and the NNSF of China (Grants No. 11404027, No. 10725521,
No. 11075020, and No. 91021021).

\end{document}